\documentclass[showpacs,aps,12pt]{revtex4}%
\usepackage{amsfonts}%
\usepackage{amsmath}%
\usepackage{graphicx}%
\usepackage{amssymb}%

\begin{document}
\title{The role of $N^*(1535)$ in $pp \to pp \phi$ and $\pi^- p \to n \phi$ reactions }
\author{ Ju-Jun Xie$^{1,4}$} \email{xiejujun@mail.ihep.ac.cn}
\author{Bing-Song Zou$^{1,2}$} \email{zoubs@mail.ihep.ac.cn}
\author{ Huan-Ching Chiang$^{1,3}$}
\affiliation{1 Institute of High Energy Physics, CAS, Beijing 100049, China\\
2 Center of Theoretical Nuclear Physics, National Laboratory of\\
Heavy Ion Accelerator, Lanzhou 730000, China\\
3 Department of Physics, South-west University, Chongqing 400071, China\\
 4 Graduate University of Chinese Academy of Sciences, Beijing
100049, China}

\begin{abstract}
The near threshold $\phi$ meson production in proton-proton and
$\pi^- p$ collisions is studied with the assumption that the
production mechanism is due to the sub-$N\phi$-threshold
$N^*(1535)$ resonance. The $\pi^0$, $\eta$ and $\rho^0$-meson
exchanges for proton-proton collisions are considered.  It is
shown that the contribution to the $pp \to pp \phi$ reaction from
the t-channel $\pi^0$ meson exchange is dominant. With a
significant $N^*(1535)N\phi$ coupling ($g^2_{N^*(1535)N \phi}/4
\pi$ = 0.13), both $pp \to pp \phi$ and $\pi^- p \to n \phi$ data
are very well reproduced. The significant coupling of the
$N^*(1535)$ resonance to $N \phi$ is compatible with previous
indications of a large $s \bar{s}$ component in the quark wave
function of the $N^*(1535)$ resonance and may be the real origin
of the significant enhancement of the $\phi$ production over the
naive OZI-rule predictions.
\end{abstract}
 \maketitle

\section{Introduction}
The meson production reaction in nucleon-nucleon collisions near
threshold has the potential to gain new information on hadron
properties~\cite{hanhart}, and the experimental database on meson
production in nucleon-nucleon collisions has expanded
significantly in recent years. On the other hand, the study of the
strangeness content of the quark wave functions of  baryons and
baryon resonances, not only in experimental side but also in
theoretical side, has been an interesting area~\cite{zouijmpa},
which is expected to provide new information on the configuration
of baryons and baryon resonances. In the naive quark model, the
nucleon and nucleon resonances have no strangeness contents,
whereas the $\phi$ meson is an ideally mixed pure $s \bar{s}$
state. From the point of view of the naive quark model the $pp \to
pp \phi$ reaction involves disconnected quark lines and is an
Okubo-Zweig-Iizuka (OZI) rule~\cite{ozi} suppressed process. The
study of  $\phi$ meson production in nucleon-nucleon reactions may
provide information on the strangeness degrees of freedom in the
nucleon or nucleon resonances and is of importance both
experimentally and theoretically.

Several years ago, the exclusive production cross section for
$\phi$ meson production in $pp$ collisions at P$_{lab}$ = 3.67
GeV/c was measured by the DISTO Collaboration~\cite{disto}, and
the preliminary result at an excess energy of 18.5 MeV above the
threshold was also published by the ANKE group~\cite{anke05}. With
these experimental information about this reaction, several
theoretical papers~\cite{sibiepja06, titov, nakaphi, kap} by using
various models were published to try to explain the experimental
data. Recently, more data at other energies  are available from
the ANKE facility~\cite{anke}. Comparing the data for the $\omega$
meson production from literature, a significant enhancement of
$\phi/\omega$ ratio of a factor 8 is found compared to predictions
based on the OZI rule. This findings require more theoretical work
to understand its origin.

It is well-known that the $N^*(1535)$ resonance couples strongly
to the $\eta$N channel. Recently, it was found that the
$N^*(1535)$ resonance has a significant coupling to $K \Lambda$ in
the analysis of the $J/\psi \to \bar{p}\Lambda K^+$ decay and the
$pp \to p\Lambda K^+$ reaction near threshold~\cite{liu}. The
analyses \cite{Mosel,Saghai} of the recent SAPHIR and CLAS $\gamma
p\to K^+\Lambda$ data \cite{ELSA,CLAS} also show a large coupling
of the $N^*(1535)$ to $K\Lambda$. In a chiral unitary coupled
channel approach it was found that the $N^*(1535)$ resonance is
dynamically generated as a pole in the second Riemann sheet with
its mass, width, and branching ratios in fair agreement with
experiments and the couplings of the $N^*(1535)$ resonance to $K
\Sigma$, $\eta N$ and $K \Lambda$ are large compared to the $\pi
N$ channel~\cite{oset}. The analyses of data on the $\eta'$
photo-production on the proton for photon energies from 1.527 to
2.227 GeV  also suggest the coupling of the $\eta' N$ channel to
the $N^*(1535)$ resonance ~\cite{dugger}.

From the naive quark model, both $\eta$ meson and $\eta'$ meson
have a $s \bar{s}$ component. It seems that the $N^*(1535)$
couples strongly to mesons with strangeness or with $s \bar{s}$
components. These phenomena indicate that there may be a
significant $s \bar{s}$ configuration in the quark wave function
of the $N^*(1535)$ resonance. So, we expect that the $N^*(1535)$
resonance may also have a significant coupling to the $\phi N$
channel.

In this paper, we assume that the productions of the $\phi$ meson
in proton-proton and $\pi^- p$ collisions are predominantly
through the excitation and decay of the sub-$\phi N$-threshold
$N^*(1535)$ resonance. By using this picture, we calculate the $pp
\to pp \phi$ and $\pi^- p \to n \phi$ reactions in the framework
of an effective lagrangian approach. By comparing with the
experimental data we find that the coupling of the $N^*(1535)$
resonance to the $\phi N$ channel needs to be somewhat larger than
its the coupling to $N\rho$ channel.  The significant coupling of
the $N^*(1535)$ resonance to $N \phi$ is compatible with previous
indications of a large $s \bar{s}$ component in the quark wave
function of the $N^*(1535)$ resonance and may be the real origin
of the significant enhancement of the $\phi$ production over the
naive OZI-rule predictions.

In the next section, we will give the formalism and ingredients in
our calculation, then numerical results and discussions are given in
Sect.3. A short summary is given in the last section.

\section{Formalism and ingredients}
\begin{figure}
\includegraphics[height=2.5in, width=3.5in]{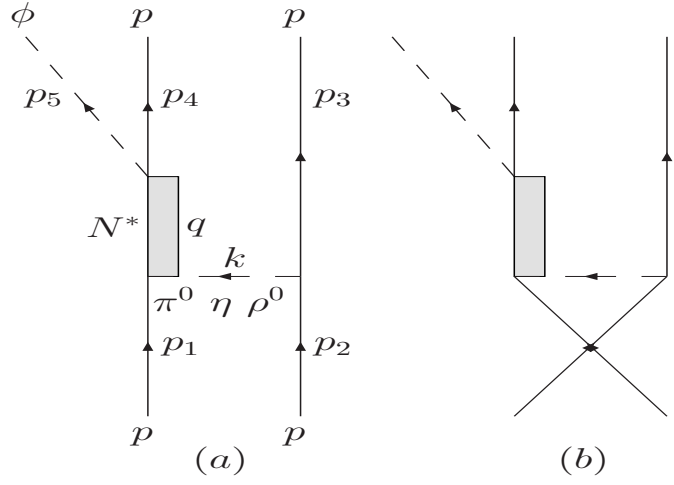}
\vspace{-0.2cm} \caption{Feynman diagrams for $pp \to pp \phi$
reaction. The diagram on the left shows the direct process, while
that on the right shows the exchange one. $p_i$ ($i$=1,2,3,4,5)
stands for the 4-momentum of the initial and final particle; $k$
and $q$ stand for the 4-momentum of exchange meson and the
intermediate resonance ($N^*(1535)$), respectively.}
\label{diagram}
\end{figure}

\begin{figure}
\includegraphics[height=1.5in, width=2.5in]{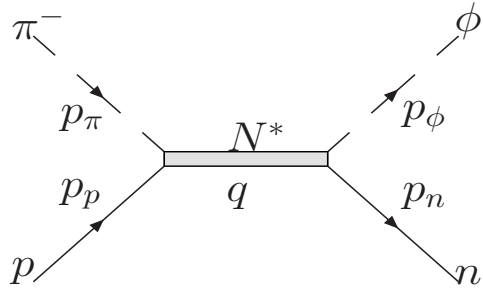}
\vspace{-0.2cm} \caption{Feynman diagram for $\pi^- p \to n \phi$
reaction. $p_{\pi}, ~ p_p, ~ p_{\phi}, ~ p_n$ and $q$ stand for
the 4-momentum of $\pi^-$, proton, $\phi$, neutron and
intermediate resonance ($N^*(1535)$), respectively.}
\label{pipdiagram}
\end{figure}

We study the $pp \to pp \phi$ and $\pi^- p \to n \phi$ reactions
near threshold in an effective Lagrangian approach. We assume that
the near threshold $\phi$ productions in proton-proton and $\pi^ -
p$ collisions are through the intermediate excitation of the
sub-$\phi N$-threshold $N^*(1535)$ resonance. The $\pi^0$, $\eta$
and $\rho^0$-meson exchanges are considered for proton-proton
collisions. The basic Feynman diagrams for the $pp \to pp \phi$
reaction and the s-channel diagram for the $\pi^- p \to n \phi$
reaction are depicted in Fig.~\ref{diagram} and
Fig.~\ref{pipdiagram}, respectively.

We use the commonly used interaction Lagrangians for $\pi NN$, $\eta
NN$ and $\rho NN$ couplings,
\begin{equation}
{\cal L}_{\pi N N}  = -i g_{\pi N N} \bar{u}_N \gamma_5 \vec\tau
\cdot \vec\pi u_N, \label{pin}
\end{equation}
\begin{equation}
{\cal L}_{\eta N N}  = -i g_{\pi N N} \bar{u}_N \gamma_5 \eta u_N,
\label{etan}
\end{equation}
\begin{equation}
{\cal L}_{\rho N N} = -g_{\rho N N}
\bar{u}_N(\gamma_{\mu}+\frac{\kappa}{2m_N} \sigma_{\mu \nu}
\partial^{\nu})\vec\tau \cdot \vec\rho^{\mu} u_N. \label{rhon}
\end{equation}

At each vertex a relevant off-shell form factor is used. In our
computation, we take the same form factors as that used in the
well-known Bonn potential model~\cite{mach}
\begin{equation}
F^{NN}_M(k^2_M)=(\frac{\Lambda^2_M-m_M^2}{\Lambda^2_M- k_M^2})^n,
\end{equation}
with n=1 for $\pi^0$ and $\eta$-meson; n=2 for $\rho^0$-meson.
$k_M$, $m_M$ and $\Lambda_M$ are the 4-momentum, mass and cut-off
parameters for the exchanged-meson ($M$), respectively. The
coupling constants  and the cutoff parameters are taken as
~\cite{mach,tsushima,sibi}: $g_{\pi NN}^2/4\pi$ = 14.4, , $g_{\rho
NN}^2/4\pi$ = 0.9, $\Lambda_{\pi}$ = $\Lambda_{\eta}$ = 1.3 GeV,
$\Lambda_{\rho}$ = 1.6 GeV, and $\kappa$ = 6.1. The value of $\eta
NN$ coupling constant is extremely uncertain, with values of
$g_{\eta NN}^2/4\pi$ between 0 and 7 being quoted in the
literature, we use $g_{\eta NN}^2/4\pi$ = 0.4 since many authors
say that it is small (see e.g. \cite{etan} and \cite{faldt}).

To calculate the invariant amplitudes of diagrams in the
Fig.~\ref{diagram} and Fig.~\ref{pipdiagram} with the $N^*(1535)$
resonance model, we also need to know the interaction Lagrangians
involving the $N^*(1535)$ resonance. In Ref.~\cite{zouprc03}, a
Lorentz covariant orbital-spin (L-S) scheme for $N^* N M$
couplings has been illustrated in detail. With this scheme, we can
easily write the effective $N^*(1535) N \pi$, $N^*(1535) N \eta$,
$N^*(1535) N \rho$ and $N^*(1535) N \phi$ couplings,
\begin{equation}
{\cal L}_{\pi N N^*}  =  ig_{N^* N\pi}\bar{u}_N
u_{N^*}+h.c.,\label{pion1}
\end{equation}
\begin{equation}
 {\cal L}_{\eta N N^*} = ig_{N^* N\eta}\bar{u}_Nu_{N^*}+h.c.,
 \label{eta1}
\end{equation}
\begin{equation}
 {\cal L}_{\rho N N^*} = ig_{N^* N\rho}\bar{u}_N \gamma_5 (\gamma_{\mu}-\frac{q_{\mu} \gamma^{\nu}
q_{\nu}}{q^2}) \varepsilon^{\mu}(p_{\rho})u_{N^*}
+h.c.,\label{rho1}
\end{equation}
\begin{equation}
 {\cal L}_{\phi N N^*} = ig_{N^* N \phi}\bar{u}_N \gamma_5 (\gamma_{\mu}-\frac{q_{\mu}
\gamma^{\nu} q_{\nu}}{q^2})
\varepsilon^\mu(p_{\phi})u_{N^*}+h.c..\label{phi1}
\end{equation}
Here $u_N$ and $u_{N^*}$ are the Rarita-Schwinger spin wave
functions for the nucleon and $N^*(1535)$ resonance;
$\varepsilon^{\mu}(p_{\rho})$ and $\varepsilon^{\mu}(p_{\phi})$
are the polarization vectors of the $\rho$ and $\phi$-meson,
respectively. It is worth noting that since the spins of the
$\rho$ meson and $\phi$ meson are 1, both  S-wave and D-wave L-S
couplings are possible for the $N^*(1535) N \rho$ and $N^*(1535) N
\phi$ interactions. It was found that  the S-wave coupling has a
significant contribution to the partial decay width of the
$N^*(1535)$ resonance compared with the D-wave
\cite{pdg2006,vrana}. In our calculation we only consider the
S-wave $N^*(1535)$ resonance couplings to $N \rho$  and neglect
the D-wave $N^*(1535)$ resonance couplings to $N \phi$ for
simplicity. The monopole form factors for $N^*(1535)$-$N$-Meson
vertexes are used,
\begin{equation}
F^{N^* N}_M(k^2_M)=\frac{\Lambda^{*2}_M-m_M^2}{\Lambda^{*2}_M-
k_M^2}, \label{sff}
\end{equation}
with $\Lambda^*_{\pi}$ = $\Lambda^*_{\eta}$ = $\Lambda^*_{\rho}$ =
1.3 GeV.


The $N^*(1535)N\pi$, $N^*(1535)N\eta$ and $N^*(1535)N\rho$
coupling constants are determined from the experimentally observed
partial decay widths of the $N^*(1535)$ resonance.  With the
effective interaction Lagrangians described by Eq.~(\ref{pion1})
and Eq.~(\ref{eta1}), the partial decay widths $\Gamma_{N^*(1535)
\to N \pi}$ and $\Gamma_{N^*(1535) \to N \eta}$ can be easily
calculated~\cite{pdg2006}. The coupling constants are related to
the partial decay widths,
\begin{eqnarray}
\Gamma_{N^*(1535) \to N \pi} &=& \frac{3 g^2_{N^*N
\pi}(m_N+E^{\pi}_N)p^{cm}_{\pi}}{4\pi M_{N^*}}, \label{1535pi}\\
\Gamma_{N^*(1535) \to N \eta} &=& \frac{g^2_{N^*N
\eta}(m_N+E^{\eta}_N)p^{cm}_{\eta}}{4\pi M_{N^*}}, \label{1535eta}
\end{eqnarray}
where
\begin{equation}
p^{cm}_{\pi / \eta}=\sqrt{\frac{(M^2_{N^*}-(m_N+m_{\pi / \eta})^2)
(M^2_{N^*}-(m_N-m_{\pi / \eta})^2)}{4M^2_{N^*}}},
\end{equation}
and
\begin{equation}
E^{\pi / \eta}_N=\sqrt{(p^{cm}_{\pi / \eta})^2+m^2_N}.
\end{equation}

For the $N^*(1535) N \rho$ coupling constant, we get it from the
partial decay width $\Gamma_{N^*(1535) \to N \rho \to N \pi \pi}$,
and the partial decay width can be evaluated from the total
invariant amplitude ${\cal M}_{N^*(1535)\to N\rho\to N\pi\pi}$ of
the $N^*(1535) \to N \rho \to N \pi \pi$ decay and a three-body
phase space integration,
\begin{eqnarray}
{\cal M}_{N^*(1535)\to N\rho\to N\pi\pi} &=& g_{\rho \pi \pi}
g_{N^*(1535) N \rho} F^{N^*N}_{\rho}(k^2_{\rho}) \bar{u}_N
(p_1,s_1) \gamma_5 (\gamma_{\mu}-\frac{q_{\mu} \gamma^{\sigma}
q_{\sigma}}{q^2}) \times \nonumber\\  &&
 G^{\mu \nu}_{\rho}(k_{\rho}) (p_2-p_3)_{\nu} u_{N^*}(q,s_{N^*}),\\
 d\Gamma_{N^*(1535) \to N
\rho \to N \pi \pi} &=& \overline{|{\cal M}_{N^*(1535)\to N\rho\to
N\pi\pi}|^2} \frac{d^3p_1}{(2 \pi)^3} \frac{m_1}{E_1}
\frac{d^3p_2}{(2 \pi)^3} \frac{1}{2 E_2} \frac{d^3p_3}{(2 \pi)^3}
\frac{1}{2 E_3} \times \nonumber\\  && (2 \pi)^4
\delta^4(q-p_1-p_2-p_3),
\end{eqnarray}
where $G^{\mu \nu}_{\rho}(k_{\rho})$ is the propagator of the $\rho$
meson with the form
\begin{equation}
G^{\mu \nu}_{\rho}(k_{\rho})=- i (\frac{g^{\mu
\nu}-k^{\mu}_{\rho}k^{\nu}_{\rho}/k^2_{\rho}}{k^2_{\rho}-m^2_{\rho}}).
\end{equation}
Here $q$ and $k_{\rho}$ are the 4-momentum of the $N^*(1535)$
resonance and the intermediate $\rho$ meson; $p_1$, $m_1$, and $E_1$
stand for the 4-momentum, mass, and energy of the nucleon; $s_1$ and
$s_{N^*}$ the spin projection of the nucleon and the $N^*(1535)$
resonance; $p_{2/3}$ and $E_{2/3}$ stand for the 4-momentum and
energy of the final two pions, respectively. In our calculation, we
use $g^2_{\rho \pi \pi}/4 \pi $ = 2.91 as the same as that used in
Ref.~\cite{lixueqian}.

There is no information for the coupling constant of the
$N^*(1535) N \phi$ vertex. We determine it from the $\pi^- p \to n
\phi$ reaction. We assume that the near threshold $\phi$
production in $\pi^- p$ collisions is through the intermediate
excitation of the sub-$\phi N$-threshold $N^*(1535)$ resonance.
Then, by comparing the theoretical total cross sections of $\pi^-
p \to n \phi$ reaction with experimental data, we can extract the
$N^*(1535) N \phi$ coupling constant.

In Fig~\ref{pipdiagram}, we show the s-channel diagram for the
$\pi^- p \to n \phi$ reaction, the intermediate excitation is a
sub-$n \phi$-threshold $N^*(1535)$ resonance. Following the
Feynman rules and with the above formula, we can obtain the
invariant amplitude $\cal A$ of the $\pi^- p \to n \phi$ reaction,
\begin{eqnarray}
{\cal A} = g_{N^* N \pi} g_{N^* N \phi} F_{N^*}(q^2)
\bar{u}(p_n,s_n) \gamma_5 (\gamma_{\mu}-\frac{q_{\mu} \gamma^{\nu}
q_{\nu}}{q^2}) \varepsilon^{\mu}(p_{\phi},s_{\phi})
G_{N^*(1535)}(q) u(p_p,s_p),
\end{eqnarray}
with $s_n$, $s_p$, $s_{\phi}$, the spin projection of the $\phi$
meson and the nucleon, respectively. The form factor for
$N^*(1535)$ resonance, $F_{N^*}(q^2)$, is taken similar as in
Refs.~\cite{Mosel, feuster}
\begin{equation}
F_{N^*}(q^2)=\frac{\Lambda^{4}}{\Lambda^{4} +
(q^2-M^2_{N^*(1535)})^2},
\end{equation}
with $\Lambda$ = 2.0 GeV. $G_{N^*(1535)}(q)$ is the propagator of
the $N^*(1535)$ resonance, which can be written in a  Breit-Wigner
form~\cite{liang},
\begin{equation}
G_{N^*(1535)}(q)=\frac{\gamma \cdot q
+M_{N^*(1535)}}{q^2-M^2_{N^*(1535)}+iM_{N^*(1535)}\Gamma_{N^*(1535)}(s)}.
\end{equation}
Here $\Gamma_{N^*(1535)}(s)$ is the energy dependent total width
of the $N^*(1535)$ resonance. According to PDG~\cite{pdg2006}, the
dominant decay channels for the $N^*(1535)$ resonance are $\pi N$
and $\eta N$, so we take
\begin{equation}
\Gamma_{N^*(1535)} (s) = \Gamma_{N^*(1535)\to N\pi}\frac{\rho_{\pi
N}(s)}{\rho_{\pi N}(M_{N^*(1535)}^2)}+\Gamma_{N^*(1535)\to
N\eta}\frac{\rho_{\eta N}(s)}{\rho_{\eta N}(M_{N^*(1535)}^2)},
\end{equation}
where $\rho_{\pi(\eta)N}(s)$ is the following two-body phase space
factor,
\begin{equation}
\rho_{\pi(\eta)N}(s)= \frac{2 p^{cm}_{\pi(\eta) N} (s)}{\sqrt{s}}
= \frac{\sqrt{(s-(m_N + m_{\pi(\eta)})^2)(s -
(m_N-m_{\pi(\eta)})^2)}}{s}.
\end{equation}

\begin{figure}
\includegraphics[height=3.5in, width=4.5in]{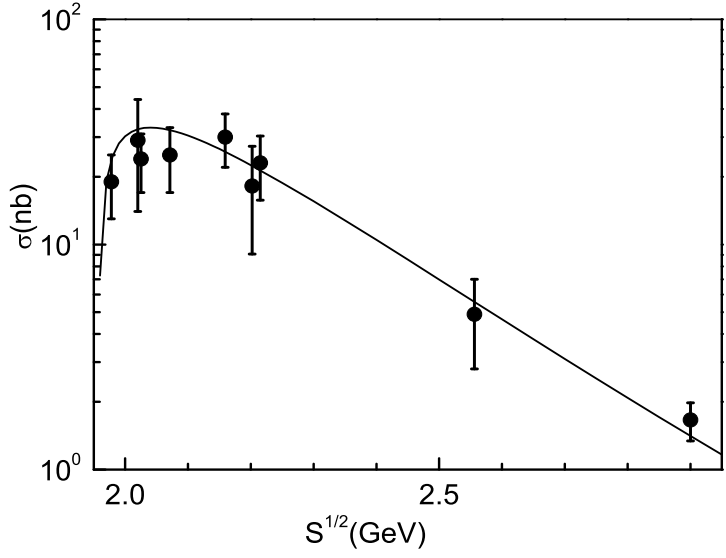}%
\vspace{-1.0cm} \caption{ Total cross sections vs the c.m. energy
$S^{\frac{1}{2}}$ for  $\pi^- p \to n
\phi$ reactions. The experimental data are from Ref.~\cite{pipdata}. } \label{piptcs}%
\end{figure}

From the amplitude, we can easily obtain the total cross sections
of the $\pi^- p \to n \phi$ reaction as functions of the excess
energies. By adjusting the $N^*(1535) N \phi$ coupling constant,
we can compare the theoretical results with the experimental data.
Theoretical results with $g^2_{N^*(1535) N \phi}/4 \pi$ = 0.13 are
compared with the experimental data in Fig.~\ref{piptcs}, we find
an excellent agreement between our results and the experimental
data. Contributions from the u-channel $N^*$ exchange and
$\rho$-meson exchange between the pion and the proton are also
checked and are found to be negligible.

With experimental mass (1535 MeV), width (150 MeV), branching
ratios of the $N^*(1535)$~\cite{pdg2006} and the total cross
sections of the $\pi^- p \to n \phi$ reaction, we obtain all the
coupling constants as listed in Table.~\ref{tab1}.

\begin{table}
\caption{\label{table} Relevant $N^*(1535)$ parameters.}
\begin{center}
\begin{tabular}{|cccc|}
\hline Decay channel  & Branching ratios & Adopted branching ratios & $g^2/4 \pi$\\
\hline $N \pi$ & 0.35-0.55 & 0.45 & 0.033 \\
       $N \eta$ & 0.45-0.60 & 0.53 & 0.28 \\
       $N \rho \to N \pi \pi $ & 0.02 $\pm$ 0.01      &0.02  & 0.10 \\
       $N \phi$ & --      &---      & 0.13 \\
\hline
\end{tabular}
\end{center} \label{tab1}
\end{table}

For the $pp \to pp \phi$ reaction, the full invariant amplitude in
our calculation is composed of three parts corresponding to the
$N^*(1535)$ resonance production from $\pi^0$, $\eta$ and
$\rho^0$-meson exchanges, respectively.
\begin{eqnarray}
{\cal M} = \sum_{i = \pi,~ \eta,~ \rho} {\cal M}_{i}. \label{amp}
\end{eqnarray}
Each amplitude can be obtained straightforwardly with the
effective couplings and following the Feynman rules. Here we give
explicitly the amplitude $\cal M_{\pi}$, as an example,
\begin{eqnarray}
{\cal M}_{\pi} & = &g_{\pi NN} g_{N^* N \pi} g_{N^* N \phi}  F^{N
N}_{\pi}(k^2_{\pi}) F^{N^* N}_{\pi}(k^2_{\pi}) F_{N^*}(q^2)
\varepsilon^{\mu}(p_{\phi},s_{\phi}) G_{\pi}(k_{\pi}) \times
\nonumber\\  && \bar{u} (p_4,s_4) \gamma_5
(\gamma_{\mu}-\frac{q_{\mu} \gamma^{\nu} q_{\nu}}{q^2})
G_{N^*(1535)}(q) u(p_1,s_1)\bar{u}(p_3,s_3) \gamma_5
 u(p_2,s_2)  \nonumber\\  &&
  + (\text {exchange term with } p_1 \leftrightarrow
p_2),
\end{eqnarray}
where $s_{\phi}$ is the spin projection of the $\phi$ meson;
$s_i~(i=1,2,3,4)$ and $p_i~(i=1,2,3,4)$ represent the spin
projection and 4-momentum of the two initial and two final
protons, respectively. $G_{\pi}(k_{\pi})$ is the pion meson
propagator,
\begin{equation}
G_{\pi}(k_{\pi})=\frac{i}{k_{\pi}^2-m^2_{\pi}}.
\end{equation}

The final-state-interaction(FSI) enhancement factor in the $^1S_0$
di-proton state are taken into account by means of the general
framework based on the Jost function formalism~\cite{gill} with
\begin{eqnarray}
|J(q)|^{-1}=\frac{k+ i \beta}{k- i \alpha}, \label{fsi}
\end{eqnarray}
where $k$ is the internal momentum of $pp$ subsystem, and the
$\alpha$ and $\beta$ are related to the scattering parameters via
\begin{eqnarray}
a=\frac{\alpha + \beta}{\alpha \beta}, ~~~~~  r=\frac{2}{\alpha +
\beta},
\end{eqnarray}
with $\alpha$ = -20.5 MeV/$c$ and $\beta$ = 166.7 MeV/$c$
\cite{sibiepja06} (i.e. $a$ = -7.82 fm and $r$ = 2.79 fm) in the
present study.

Then the calculations of the differential and total cross sections
are straightforward,
\begin{eqnarray}
d\sigma (pp\to pp \phi)=\frac{1}{4}\frac{m^2_p}{F} \sum_{s_i}
\sum_{s_f} |{\cal M}|^2\frac{m_p d^{3} p_{3}}{E_{3}} \frac{m_p
d^{3} p_4}{E_4} \frac{d^{3} p_5}{2 E_5} \delta^4
(p_{1}+p_{2}-p_{3}-p_{4}-p_5), \label{eqcs}
\end{eqnarray}
with the flux factor
\begin{eqnarray}
F=(2 \pi)^5\sqrt{(p_1\cdot p_2)^2-m^4_p}~. \label{eqff}
\end{eqnarray}

Since the relative phases among different meson exchanges in the
amplitude of Eq.~(\ref{amp}) are not known, the interference terms
are ignored in our concrete calculations.

\section{Numerical results and discussions}
\begin{figure}
\includegraphics[height=3.0in, width=4.2in]{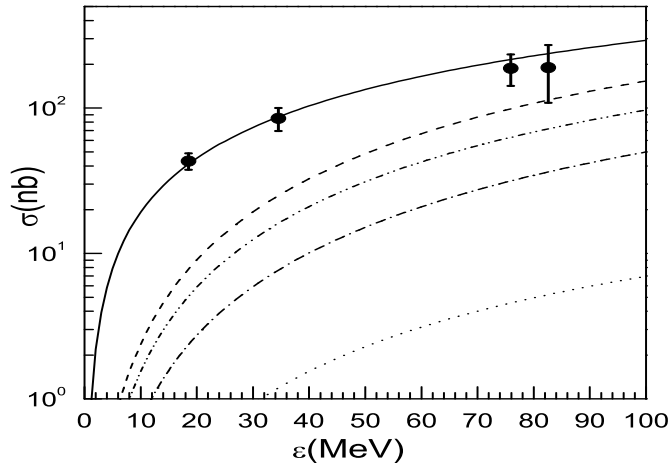}%
\vspace{-1.0cm} \caption{Total cross sections vs excess energies
($\varepsilon$) for the $pp \to pp\phi$ reaction from present
calculation are compared with experimental data~\cite{disto,
anke}. The double dotted-dashed, dotted, dashed-dotted and dashed
curves stand for contributions from $\pi^0$, $\eta$,
$\rho^0$-meson exchanges and their simple sum,
respectively. Solid line corresponds to the results with the $^1S_0$ $pp$ FSI.} \label{tcs}%
\end{figure}

With the formalism and ingredients given above, the total cross
section versus excess energy $\varepsilon$ for the $pp \to pp
\phi$ reaction is calculated by using a Monte Carlo multi-particle
phase space integration program. It is known that the
near-threshold production of the $\eta$ meson in $pp \to pp \eta$
reaction is thought to occur predominantly via the excitation of
the $N^*(1535)$ resonance. However, the excitation mechanism of
the $N^*(1535)$ resonance in proton-proton collisions is currently
still debated. For example, Batini\'c et al.~\cite{bati} and
Nakayama~\cite{nakaeta} have found that the $\pi$ and $\eta$-meson
exchanges between two protons play dominant roles for the
excitation of the $N^*(1535)$ resonance. However, Gedalin et
al.~\cite{geda} and F\"aldt and Wilkin~\cite{faldt} have found
that the $\rho$-meson exchange is the dominant excitation
mechanism of the $N^*(1535)$ resonance. Here the $\pi^0$, $\eta$
and $\rho^0$-meson exchanges for $N^*(1535)$ excitation are all
considered. By using the formalism and ingredients described in
past section we first study the roles of different meson exchanges
in the $pp \to pp \phi$ reaction. Our calculated results are shown
in Fig.~\ref{tcs} together with the experimental data. The
double-dotted-dashed, dotted and dashed-dotted curves stand for
contributions without the $pp$ final-state-interaction (FSI) from
$\pi^0$, $\eta$ and $\rho^0$-meson exchanges, respectively. A
simple summation of them are shown by the dashed line. One can see
that the contribution from the t-channel $\pi^0$ meson exchange is
dominant to the $pp \to pp \phi$ reaction in our model. The
$\rho^0$-meson exchange has a significant contribution to this
reaction, while the contribution from the $\eta$-meson exchange is
negligible.

From Fig.~\ref{tcs} we can see that our theoretical result without
the $pp$ FSI agrees well with the experimental data at excess energy
$\varepsilon$ = 83.0 MeV. However, at lower excess energies such as:
$\varepsilon$ = 18.5, 34.5 MeV, the calculated total cross sections
are lower than the data by a factor of more than 4. It is known that
the proton-proton FSI plays an important role for the near threshold
meson production in proton-proton collisions. We also include the
effect of the $^1S_0$ $pp$ FSI by using the Jost-function
method~\cite{gill} in our calculation, the results are shown in
Fig.~\ref{tcs} by the solid line which can reproduce the ANKE total
cross section data well.

\begin{figure}
\includegraphics[height=3.8in, width=4.5in]{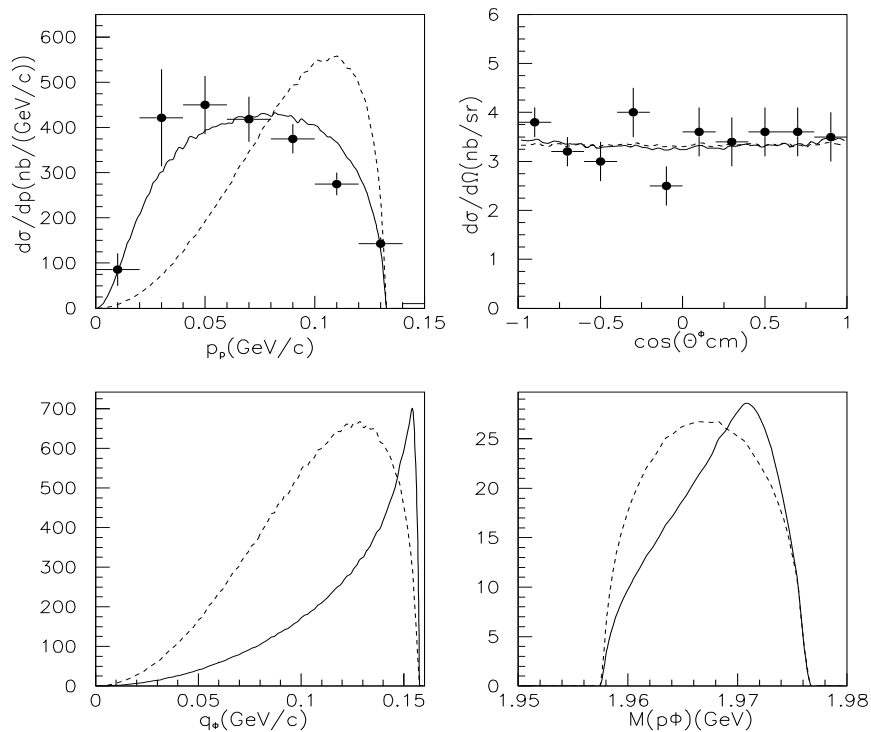}%
\vspace{-0.6cm} \caption{Differential cross sections (solid lines)
for the $pp \to pp \phi$ reaction at the excess energy
$\varepsilon$ = 18.5 MeV compared with the ANKE data~\cite{anke05}
and phase space distribution (dashed lines). The upper left panel
is the momentum distribution of the outgoing proton. The upper
right panel is the angular distribution of the $\phi$ meson in the
total c.m. frame; The lower left panel is the distribution of the
c.m. momentum of the $\phi$ meson; The lower right panel is the
invariant mass spectrum of the outgoing proton and the $\phi$
meson. }
\label{phi185}%
\end{figure}

The momentum, angular distributions of the $\phi$ meson and the $p
\phi$ invariant mass spectrum for the $pp \to pp \phi$ reaction at
excess energy $\varepsilon$ = 18.5 MeV and 83.0 MeV are also
calculated. In Fig.~\ref{phi185} we present our calculated results
at excess energy $\varepsilon$ = 18.5 MeV together with
experimental data from the ANKE group. Differential cross sections
as a function of the c.m. momentum of the outgoing proton are
presented in the upper left panel. The upper right panel is the
angular distribution of the $\phi$ meson in the total
proton-proton c.m. frame.  The dashed lines are pure phase space
distributions, while, the solid lines are full calculations from
our model with the $^1S_0$ $pp$ FSI enhancement factor. By
comparing with the data, we find that the $pp$ FSI plays an
important role. Our model can explain the experimental data well.
In the lower part of FIG.~\ref{phi185} the momentum distribution
of the $\phi$ meson and the invariant mass spectrum of the
outgoing proton and the $\phi$ meson are shown.

\begin{figure}
\includegraphics[height=3.8in, width=4.5in]{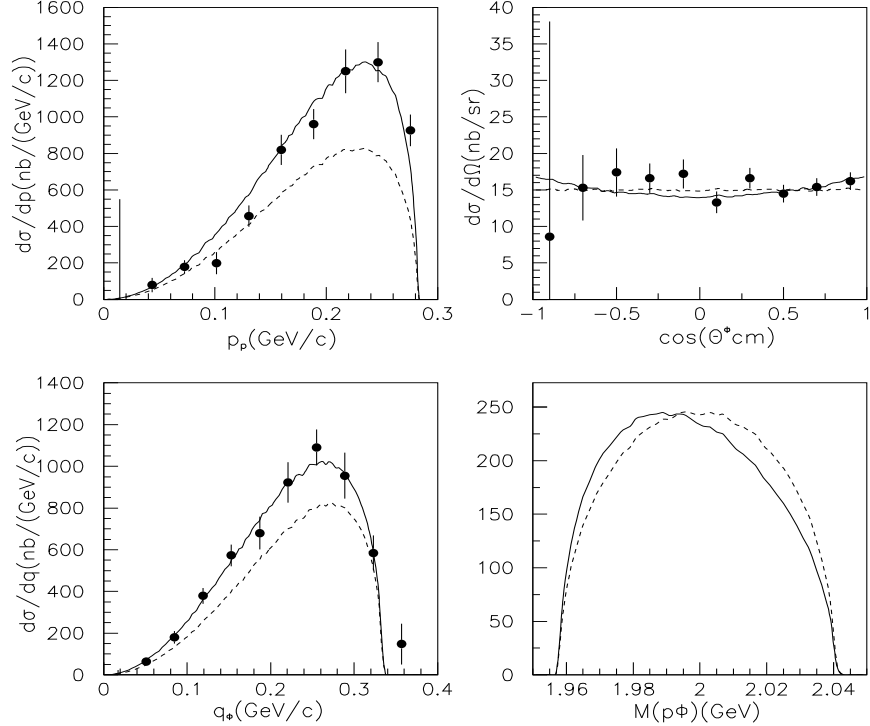}%
\vspace{-0.6cm} \caption{Differential cross sections for the $pp
\to pp \phi$ reaction at the excess energy $\varepsilon$ = 83 MeV
compared with the DISTO data~\cite{disto}. The dashed line
reflects pure phase space, while the solid lines, which includes
the amplitudes but without the $pp$ FSI.}
\label{phi83}%
\end{figure}

In Fig.~\ref{phi83}, we present our calculated differential
distributions at excess energy $\varepsilon$ = 83.0 MeV  together
with experimental data from the DISTO group. From our calculation
we find that there is no need to consider the $pp$ FSI at this
energy. An excellent agreement between our model calculation  and
the experimental data both in shapes and magnitudes can be
achieved without taking the $pp$ FSI into account. This is
consistent with ANKE findings at $\varepsilon$ = 75.9 MeV. The
phenomena may suggest that at excess energy about 80 MeV the
contribution from $pp$ higher partial waves has already overtaken
the $^1S_0$ partial wave as the dominant contribution and the FSI
becomes not important.

In our calculation we only include the contribution of the
$N^*(1535)$ in the intermediate state. In previous calculations
\cite{sibiepja06, titov, nakaphi, kap}, the $\pi p\to\phi N$
through t-channel $\rho$ exchange and/or sub-threshold nucleon
pole contributions are assumed to be dominant. However these
contributions are very sensitive to the choice of off-shell form
factors for the t-channel $\rho$ exchange and the $g_{NN\phi}$
couplings and can be reduced by orders of magnitude within the
uncertainties of these ingredients. Considering the ample evidence
for large coupling of the $N^*(1535)$ to the strangeness
\cite{liu,Mosel,Saghai,oset,Riska} and the $N^*(1535)$ resonance
is closer than the nucleon pole to the $\phi N$ threshold, it is
more likely that the $N^*(1535)$ plays dominant role for near
threshold $\phi$ production from $\pi p$ and $pp$ collisions
instead of the nucleon pole or the OZI suppressed $\phi\rho\pi$
coupling. Our calculation with the $N^*(1535)$ domination
reproduces energy dependence of the $\pi^-p\to\phi n$ and $pp\to
pp\phi$ cross sections better than previous calculations. The
significant coupling of the $N^*(1535)$ resonance to $N \phi$ may
be the real origin of the significant enhancement of the $\phi$
production from $\pi p$ and $pp$ reactions over the naive OZI-rule
predictions. This makes it difficult to extract the properties of
the strangeness in the nucleon from these reactions proposed by
J.Ellis et al \cite{Ellis}. There are also some suggestions
\cite{gao,zhang} for possible existence of an $N\phi$ bound state
just below the $N\phi$ threshold. However, contribution of such
bound state with width less than 100 MeV will give a much sharper
dropping structure for the $\pi^-p\to\phi n$ cross section at
energies near threshold. If such $N\phi$ bound state does exist,
it should have weak coupling to $\pi N$ and only gives small
contribution to the $\pi^-p\to\phi n$ reaction.

\section{Conclusions }
In this paper, the  near threshold $\phi$ meson productions in
proton-proton and $\pi^- p$ collisions are studied with an
effective Lagrangian approach. We assume that the production
mechanism is due to the excitation of the sub-$N\phi$-threshold
$N^*(1535)$ resonance following $\pi^0$, $\eta$ and $\rho^0$-meson
exchanges between two protons. $\pi^0 NN$, $\eta NN$ and $\rho^0
NN$ coupling constants (except $g_{\eta NN}$) and form factors are
taken from the Bonn potential model. $N^*(1535) N \pi^0$,
$N^*(1535) N \eta$ and $N^*(1535) N \rho^0$ coupling constants are
determined from the partial decay widths of the $N^*(1535)$
resonance. The $N^*(1535) N \phi$ coupling constant is deduced
from a fit to the experimental total cross sections of the $\pi^-
p \to n \phi$ reaction near threshold with the $N^*(1535)$
resonance model. We find that the $N^*(1535)$ resonance has a
significant coupling to $N \phi$ ($g^2_{N^*(1535)N \phi}/4 \pi$ =
0.13).

The total reaction cross sections and differential distributions of
the near threshold $pp \to pp \phi$ reaction are calculated with the
$N^*(1535)$ resonance model without adjustable parameter. Our
theoretical calculation agrees quite well with experiments near
threshold. We find that the contribution from the t-channel $\pi^0$
meson exchange is dominant to the $pp \to pp \phi$ reaction.

The significant coupling of the  $N^*(1535)$ resonance to the
$\phi N$ channel together with the earlier findings of large
couplings of the $N^*(1535)$ resonance to the $\eta N$, $\eta' N$
and $K \Lambda$ channels \cite{liu,Mosel,Saghai,dugger,pdg2006}
gives a coherent picture that there is a large component of
strangeness in the $N^*(1535)$ resonance as expected by various
theoretical approaches \cite{liu,oset,Riska,zhusl}. It also gives
a natural explanation for the significant enhancement of the
$\phi$ production from $\pi p$ and $pp$ reactions over the naive
OZI-rule predictions. For a better understanding of the dynamics
of these reactions, more experimental data at other excess
energies with Dalitz plots and angular distributions are desired.

\bigskip
\noindent
{\bf Acknowledgement}\\
We would like to thank Feng-kun Guo, Bo-chao Liu, Colin Wilkin,
Feng-quan Wu and Hai-qing Zhou for useful discussions. This work
is partly supported by the National Natural Science Foundation of
China under grants Nos. 10435080, 10521003 and by the Chinese
Academy of Sciences under project No. KJCX3-SYW-N2.

\end{document}